\documentclass[twocolumn,amsmath,amssymb]{revtex4}
\usepackage{graphicx}
\usepackage{bm}
\begin{document}
\title{
Oscillations of magnetoresistance of 2DEG in a weak magnetic field under
microwave irradiation
}
\author{A. E. Patrakov}
\author{I. I. Lyapilin}
\email{Lyapilin@imp.uran.ru}
\affiliation{Institute of Metal Physics, UD of RAS, Yekaterinburg, Russia}
\begin{abstract}
Under microwave irradiation, in 2D electron systems with high
filling factors oscillations of longitudinal magnetoresistance
appear in the range of magnetic fields where ordinary SdH
oscillations are suppressed. An unusual beat-like behaviour of
these oscillations in weak magnetic fields ($B \le 0.02$~T) was
attributed to the zero spin splitting previously. We propose an
alternate explanation of this beat-like structure on the basis of
the Boltzmann kinetic equation, without any reference to the
spin-orbit interaction.
\end{abstract}

\maketitle

The interest to theoretical investigations of non-linear transport
phenomena in two-dimensional electron systems (2DES) has grown
substantially due to new experimental results obtained in very
clean 2DES samples. It has been discovered independently by two
experimental groups \cite{Zudov, Mani} that the resistance of
two-dimensional high-mobility electron gas in $GaAs/AlGaAs$
heterostructures reveals a series of new features in its
dependence on magnetic field, temperature, radiation power, etc.
Under microwave irradiation, in 2DES with high filling factors
oscillations of longitudinal magnetoresistance appear in the range
of magnetic fields where ordinary SdH oscillations are suppressed.

\begin{figure}
\center\includegraphics{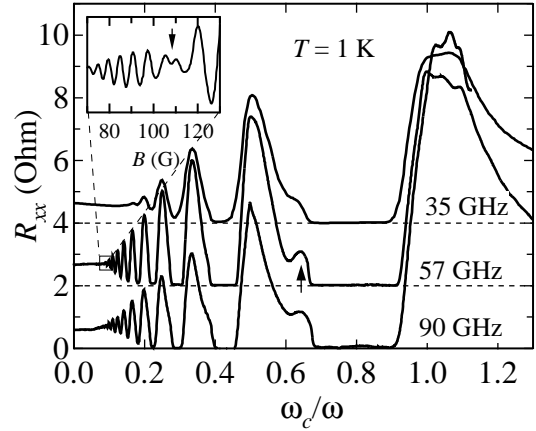} \caption{$R_{xx}$ ($T$ = 1
K) in the presence of microwave vs $\omega_c/\omega$ (for clarity, plots
have been offset vertically). The inset shows with magnification
the part of the plot, corresponding to the frequency of 57 GHz and weak
magnetic fields. At $B \lesssim 0.1$~kG the ``beat'' can be seen.
(This figure was taken from \cite{Zudov})} \label{fig:Zudov1b}
\end{figure}

\begin{figure}
\center\includegraphics{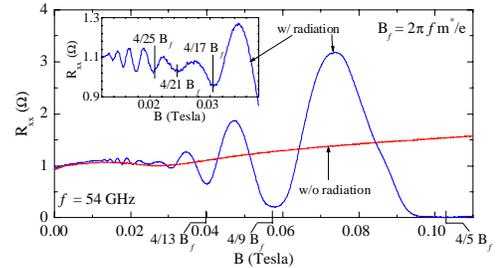} \caption{The resistance
$R_{xx}$ of 2DEG in the $GaAs/AlGaAs$ heterostructure with and without
microwave irradiation. The ``beat'' visible in resistance oscillations
is expanded in the inset. (This figure was taken from \cite{ManiCM})}
\label{fig:Mani1a}
\end{figure}

Among the new effects, an unusual behavior of longitudinal
magnetoresistance in weak magnetic fields \linebreak $B\leq0.02$~T
is worth mentioning. E.g., in experiments by Zudov \cite{Zudov}
performed in $GaAs/AlGaAs$ heterostructures (with electron
mobility $\mu=25\times 10^6$~cm$^2$/Vs and electron density
$n_e=3.5\times 10^{11}$~cm$^{-2}$) vanishing of amplitude of
magnetoresistance oscillations has been observed at $B=110$~T.
Upon further lowering of the magnetic field, the amplitude of
oscillations grows again (see the inset in Fig.
\ref{fig:Zudov1b}). This feature of magnetoresistance has been
interpreted as a beat. The analogous effect has been also observed
by Mani \cite{ManiCM} in $GaAs/AlGaAs$ heterostructures (with
electron mobility $\mu=15\times 10^6$~cm$^2$/Vs and density
$n_e=3\times 10^{11}$~cm$^{-2}$). Vanishing of the amplitude of
oscillations has been observed in the vicinity of $B=240$~G (see
Fig. \ref{fig:Mani1a}). According to \cite{ManiCM}, the magnetic
field at which the amplitude of magnetoresistance oscillations
vanishes does not depend on microwave radiation frequency for the
same sample, although the effect itself is visible only in the
presence of radiation. The observed ``beats'' were associated by
Mani with ``zero-spin splitting'' resulting from the spin-orbit
interaction \cite{Rashba, Lommer}.

In this paper we will show that the nature of the observed ``beat'' can be
understood in a simple framework without the need for any spin-orbit
interaction. First, let's mention important conditions of the experiments
discussed above: the effect is seen at
$$\hbar/\tau\ll T \simeq \hbar\omega_c \leq \hbar\omega \ll
\zeta$$ It follows that the effect has a quasiclassical nature.
Here $\tau$ is the momentum relaxation time, $\omega < \omega_c$
is the radiation and cyclotron frequencies respectively, $\zeta$
denotes the Fermi energy, $T$ is the temperature expressed in
units of energy.

Since the ``beat'' expresses itself in the classical range of
magnetic fields, we will start from the Boltzmann kinetic equation.
Taking into account that the electrons are driven from the
equilibrium both by external dc electric field and the electric
field of radiation, we have:
\begin{equation}
\left.\frac{\partial f(\bm p)}{\partial t}\right|_{E_{ac}} -e \bm
E_{dc} \frac{\partial f(\bm p)}{\partial \bm p} -\frac{e}{m c}\left[
\bm p \times  \bm H \right] \frac{\partial f(\bm p)}{\partial \bm
p} = I_{st}[f(\bm p)].
\end{equation}
Here $E_{dc}$ is the strength of the external dc field.
The right hand side is the collision integral. We assume that the main
electron scattering mechanism is the elastic scattering on impurities:
\begin{equation}
I_{st}[f(\bm p)] = - \frac{f(\bm p)-f_0(\bm p)}{\tau},
\end{equation}
where $f_0(\bm p)$ is the equilibrium distribution function, $\tau$ is the
relaxation time.
$\left.\frac{\partial f(\bm
p)}{\partial t}\right|_{E_{ac}}$
is the rate of distribution function change due to transitions between Landau
levels caused by microwave radiation:
\begin{equation}
\left.\frac{\partial f(\bm p)}{\partial t}\right|_{E_{ac}}=
\sum_{\bm p'} w_{\bm p \bm p'} (f(\bm p') - f(\bm p)),
\end{equation}
where $w_{\bm p \bm p'}$ is the probability of the electron transition
from the state with the kinetic momentum $\bm p$ to the state with the kinetic momentum $\bm p'$ in a unit of time due to microwave radiation.

The non-equilibrium distribution function can be represented in the following
form:
\begin{equation}\label{fdist}
f(\bm p) = f_0(\varepsilon(\bm p)) + \bm p \bm g(\varepsilon(\bm
p)),
\end{equation}
where $f_0(\varepsilon(\bm p))$ is the equilibrium distribution function
and $\bm g(\varepsilon(\bm p))$ is an unknown function that depends only on
electron energy.

Inserting \eqref{fdist} into the kinetic equation and linearizing upon $E_{dc}$,
we obtain:
\begin{multline}
\sum_{\bm p'} w_{\bm p \bm p'} (f_0(\varepsilon') -
f_0(\varepsilon)
+ \bm p' \bm g(\varepsilon') - \bm p \bm g(\varepsilon))-\\
\frac{e}{m} \bm E_{dc} \bm p \frac{\partial
f_0(\varepsilon)}{\partial \varepsilon} -\frac{e}{m c}\left[ \bm p
\times  \bm H \right]\bm g(\varepsilon) = -\frac{\bm p \bm
g(\varepsilon)}{\tau}.
\end{multline}

The main effect of the ac electric field amounts to changes in electron
energy (and not its momentum). Taking all simplifications mentioned into
account, we can express the kinetic equation in the following form:
\begin{multline}
\label{tosolve} \sum_{\bm p'} w_{\bm p \bm p'} (\bm
g(\varepsilon') - \bm g(\varepsilon)) -\frac{e}{m} \bm E_{dc}
\frac{\partial f_0(\varepsilon)}{\partial \varepsilon} -\\
\frac{e}{m c}\left[ \bm H \times \bm g(\varepsilon) \right] +\frac{\bm
g(\varepsilon)}{\tau}  = 0.
\end{multline}

We shall search for a solution of \eqref{tosolve} in the form of a power
series on $w_{\bm p \bm p'}$. Up to the linear terms, we have:
\begin{multline}
\label{solv} \bm g(\varepsilon) = \bm g_0(\varepsilon) - \left(
\frac{\tau}{1+\omega_c^2 \tau^2} \right)^2 \frac{e}{m}
((1-\omega_c^2\tau^2) \bm E_{dc} +
\\
2 \tau[\bm \omega_c \times \bm E_{dc}])\times
\sum_{\bm p'}w_{\bm p \bm p'} \left(\frac{\partial
f_0(\varepsilon')}{\partial \varepsilon'} -\frac{\partial
f_0(\varepsilon)}{\partial \varepsilon}\right),
\end{multline}
where
\begin{equation}
\label{iter1solved} \bm
g_0=\frac{\tau}{1+\omega_c^2\tau^2}\frac{e}{m} (\bm E_{dc} + \tau[\bm
\omega_c \times \bm E_{dc}]) \frac{\partial
f_0(\varepsilon)}{\partial \varepsilon}.
\end{equation}
Here $\bm \omega_c=(e \bm H)/(m c)$ and it has been taken into account
that $\bm E_{dc} \perp \bm H$.

Performing simple transformations of \eqref{solv}, we obtain:
\begin{multline}
\label{soldg} \bm g(\varepsilon) = \bm g_0(\varepsilon) - \left(
\frac{\tau}{1+\omega_c^2 \tau^2} \right)^2 \frac{e}{m}
((1-\omega_c^2\tau^2) \bm E_{dc} +\\
2 \tau[\bm \omega_c \times \bm E_{dc}])
\times \sum_\pm \rho(\varepsilon\pm\hbar\omega) w_\pm
\left(\frac{\partial f_0(\varepsilon\pm\hbar\omega)}{\partial
\varepsilon} -\frac{\partial f_0(\varepsilon)}{\partial
\varepsilon}\right),
\end{multline}
where  $w_\pm \propto E^2_{ac} \ell^2 \left| \langle n \pm 1 | x | n
\rangle \right| ^2 \propto E^2_{ac} n / \omega_c \propto E^2_{ac}
\varepsilon_n / \omega_c^2$, $E_{ac}$ is the amplitude of the ac electric
field with the frequency of
$\omega$, $\rho(\varepsilon)$ is the density of states, and $\ell=\sqrt{\hbar
/ (m \omega_c)}$ is the magnetic length.

Using the correction to the distribution function calculated above, we find
the current density caused by radiation:
\begin{equation}
\label{jp} \bm j = \frac{2 e}{(2 \pi \hbar)^2 \rho_0 m}
\int_0^{2\pi} \int_0^\infty p' d p' d \varphi \rho(\varepsilon')
\bm p' (\bm p' \bm g(\varepsilon')),
\end{equation}
where $\rho_0=m/(2 \pi\hbar^2)$ is the density of states without the magnetic
field for one spin direction.

Inserting the explicit form of $\bm g$ from \eqref{soldg} into \eqref{jp}, we
obtain for diagonal components of the conductivity tensor:
\begin{multline}
\sigma_{xx}^{ph} = - \frac{2 e^2}{m} \frac{\tau^2 (\omega_c^2\tau^2 -
1)}{(1+\omega_c^2 \tau^2)^2} \times \\
 \sum_\pm
\int_0^\infty d\varepsilon \ \varepsilon\  w_\pm\
\rho(\varepsilon) \rho(\varepsilon\pm\hbar\omega)
\left(\frac{\partial f_0(\varepsilon\pm\hbar\omega)}{\partial
\varepsilon} -\frac{\partial f_0(\varepsilon)}{\partial
\varepsilon}\right)
\end{multline}

Taking the energy integral under the assumption of strong degeneracy of
electrons, we have:
\begin{multline}
\sigma_{xx}^{ph}\propto    - \frac{\tau^2
(\omega_c^2\tau^2-1)}{\omega_c^2(1+\omega_c^2 \tau^2)^2}
\rho(\zeta) \times\\
\sum_\pm \left( \zeta^2 \rho(\zeta\pm\hbar\omega) -
(\zeta \mp \hbar\omega)^2 \rho(\zeta \mp \hbar\omega) \right),
\end{multline}
where $\zeta $ is the Fermi energy.
Using an explicit expression for the
density of states
\cite{Ando}
\begin{eqnarray}
\rho(\varepsilon)=\rho_0(1-\delta\cos(\frac{2\pi\varepsilon}{\hbar\omega_c})),
\nonumber\\
\delta=2 e^{-\pi/(\omega_c\tau_f)} \ll 1
\end{eqnarray}
($\tau_f$ is a single-particle lifetime without magnetic field)
we obtain the following expression for diagonal photoconductivity:
\begin{equation}\label{full}
\sigma_{xx}^{ph}-\sigma_{xx}^{ph,\ SdH} \propto
\frac{\omega^2\tau^2 (\omega_c^2\tau^2-1)}{\omega_c^2(1+\omega_c^2
\tau^2)^2} \delta^2 \cos\frac{2\pi\omega}{\omega_c}
\end{equation}
It follows from \eqref{full} that photon-assisted conductivity as
a function of inverse magnetic field gives rise to slow
oscillations: $\propto\cos(2\pi \omega/\omega_c)$, whereas
usual SdH oscillations denoted as $\sigma_{xx}^{ph,\ SdH}$ contain factors
like $\cos (2\pi\zeta/\hbar\omega_c)$, are fast and their amplitude decreases
rapidly when increasing temperature.

\begin{figure}
\center\includegraphics[width=7cm]{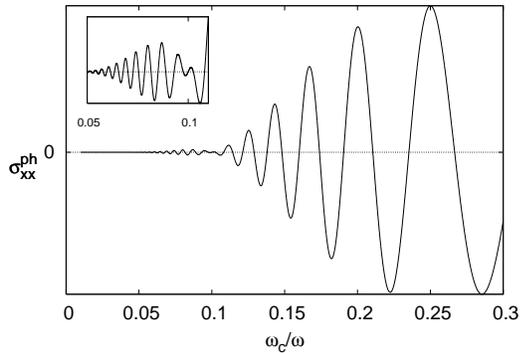}
\caption{Oscillations of 2DES
photoconductivity in the magnetic field,
$\omega\tau=10$}\label{fig:theory}
\end{figure}

The dependency of photoconductivity upon $\omega_c/\omega$, calculated
according to Eq. \eqref{full}, is plotted on Fig. \ref{fig:theory}.

It also follows from Eq. \eqref{full} that at $\omega_c \tau =
1$ the amplitude of radiation-induced magnetoresistance oscillations vanishes.
This is the very fact responsible for the ``beat'' observed in experiments
\cite{Zudov,Mani}.
Note that the observation of the node of the ``beat'' in stronger magnetic
fields in experiments \cite{Mani} as compared to \cite{Zudov} is (as one can
easily verify) a direct consequence of the lower electron mobility in
experiments \cite{Mani}.

\end{document}